\input harvmac
\sequentialequations
%%%%%%%%%%Definitions%%%%%%%%%
\def\cqg#1{{\it Class. Quan. Grav.,} {\bf #1},}
\def\np#1{{\it Nucl. Phys.,} {\bf B#1},}
\def\prl#1{{\it Phys. Rev. Lett.,} {\bf #1},}

\def\plb#1{{\it Phys. Lett.,} {\bf B#1},}

\def\pr#1{{\it Phys. Rev.,} {\bf #1},}

\def\frac#1#2{\hbox{$#1\over #2$}}

\def\tilde{\widetilde}

\def\ce{{\cal E}}
   %%% supercharge
\def\mybullet#1{\smallskip
\noindent$\bullet$ {\sl #1}\par\nobreak\noindent}

\xdef\mysecsym{2.}

\global\newcount\subsecno \global\subsecno=0
\def\subsec#1{\global\advance\subsecno by1
\message{(\mysecsym\the\subsecno. #1)}
\ifnum\lastpenalty>9000\else\bigbreak\fi
\noindent{\it\mysecsym\the\subsecno. #1}\writetoca{\string\quad 
{\mysecsym\the\subsecno.} {#1}}\par\nobreak\medskip\nobreak}

\def\appendix{\global\meqno=1\global\subsecno=0\xdef\secsym{\hbox{A.}}
\bigbreak\bigskip\noindent{\bf Appendix  }
\writetoca{Appendix  }\par\nobreak\medskip\nobreak}

%%%%%%%%%%%%%%references%%%%%%%%%%

\lref\aezero{P.~C.~Aichelburg, and F.~Embacher,
``Exact superpartners of N=2 supergravity solitons,'' Phys. Rev. D34 (1986) 3006.}

\lref\aeone{P.~C.~Aichelburg, and F.~Embacher, ``Supergravity Solitons I: General Framework,''
Phys. Rev. D37 (1988) 338.}

\lref\aetwo{P.~C.~Aichelburg, and F.~Embacher, ``Supergravity Solitons II: the Free Case,''
Phys. Rev. D37 (1988) 911.}

\lref\aethree{P.~C.~Aichelburg, and F.~Embacher, ``Supergravity Solitons III: the Background 
Problem,'' Phys.Rev. D37 (1988) 1436.}

\lref\aefour{P.~C.~Aichelburg, and F.~Embacher,
``Supergravity Solitons IV: Effective Soliton Interaction,''
Phys. Rev. D37 (1988) 2132.}

\lref\duff{M.~J.~Duff, and K.~S.~Stelle, 
``Multimembrane Solutions of D=11 Supergravity," Phys. Lett. B273 (1991) 113.}

\lref\duffone{M.~J.~Duff,  J.~T.~Liu, and J.~Rahmfeld, 
``Dipole Moments of Black Holes and String States,''
Nucl. Phys. B494 (1996) 161, hep-th/9612015.}

\lref\dufftwo{M.~J.~Duff,  J.~T.~Liu, and J.~Rahmfeld, 
``g=1 for  Dirichlet 0-branes,'' Nucl. Phys. B524 (1998) 129,
hep-th/9801072.}

\lref\jackiw{R.~Jackiw and C.~Rebbi, 
``Solitons with fermion number $\half$,'' Phys.Rev. D13 (1976) 3398.}

\lref\bkt{V.~Balasubramanian, D.~Kastor and J.~Traschen, to appear.}

\lref\maction{E.~Bergshoeff, E.~Sezgin and P.~K.~Townsend, 
``Properties of the Eleven-Dimensional Supermembrane Theory,''
Annals Phys. 185 (1988) 330.}

\lref\harvey{J.~Harvey, ``Magnetic Monopoles, Duality and Supersymmetry'', 
in Trieste HEP Cosmology 1995, hep-th/9603086 }

\lref\cremmer{E.~Cremmer, B.~Julia, and J.~Scherk 1978, 
``Supergravity Theory in Eleven-dimensions,'' \plb{76} 409-412.}

\lref\gkt{J.~Gauntlett, D.~Kastor, and J.~Traschen 1996, 
``Overlapping Branes in M-Theory,'' \np{478} 544-560, 
hep-th/9604179 and references therein.}

\lref\dewit{B.~de Wit, K.~Peeters, and J.~Plefka, ``Superspace Geometry 
for Supermembrane Backgrounds,'' hep-th/9803209.}

\lref\iwp{Z.~Perj\'es, ``Solutions of the Coupled Einstein 
Maxwell Equations 
Representing the Fields of Spinning Sources,'' Phys. Rev. Lett. 27 (1971) 1668.
\hfill\break
W.~Israel, and G.~A.~Wilson, ``A Class of Stationary Electromagnetic 
Vacuum Fields,'' 
J. Math. Phys. 13 (1972) 865.}

\lref\wald{R.~M.~Wald, 
``Gravitational Spin Interaction,'' Phys. Rev. D6 (1972) 406.}

\lref\youm{M.~Cveti\v c, and D.~Youm,
``Rotating Intersecting M-branes,''
Nucl. Phys. B499 (1997) 253, hep-th/9612229.}

\lref\schwinger{W.~Rarita, and J.~Schwinger 1941,
``On a Theory of Particles with Half Integral Spin,'' \pr{60} 61-61.}

\lref\gibbons{See for example
H.~L\"u, and C.~N.~Pope 1996, ``p-brane Solitons in Maximal
Supergravities," \np{465} 127-156, hep-th/9512012.\hfil\break
 G.~W.~Gibbons, G.~T.~Horowitz, and
P.~K.~Townsend 1995, 
``Higher-dimensional resolution of dilatonic black hole singularities,"
\cqg{12} 297-318, hep-th/9410073.}

\lref\ruiz{A.~Dabholkar, G.~Gibbons, J.~A.~Harvey, and F.~R.~Ruiz 1990,
``SUPERSTRINGs AND SOLITONS,''
\np{340} 33-55.}

\lref\townsend{J.~A.~de Azcarraga, J.~P.~Gauntlett,
J.~M.~Izquierdo, and P.~K~Townsend 1989, ``Topological Extensions of the 
Supersymmetry Algebra for Extended Objects," \prl{63} 2443-2447.}

\lref\gueven{R.~G\"uven ``Black P-brane Solutions of D=11
Supergravity Theory,'' Phys. Lett.  {276B} (1982) 49.}

\lref\deser{C.~Teitelboim 1977, ``Surface Integrals as Symmetry 
Generators in Supergravity Theory,'' \plb{69} 240-244.\hfill\break
S.~Deser, and C.~Teitelboim 1977,
``Supergravity Has Positive Energy,''
\prl{39} 249-252.
}

\lref\teit{C.~Teitelboim,  ``Surface Integrals as Symmetry 
Generators in Supergravity Theory,'' Phys. Lett. 69B (1977) 240-244.}

\lref\harveyspin{J.~A.~Harvey, ``Spin Dependence of D0-Brane Interactions,''
Nucl. Phys. Proc. Suppl. 68 (1998) 113, hep-th/9706039.}

\lref\krausspin{P.~Kraus, ``Spin Orbit Interaction from Matrix Theory'',
Phys. Lett. B419 (1998) 73-78, hep-th/9709199.} 

\lref\morales{J.~F.~Morales, C.~Scrucca and M.~Serone,
``A Note on Supersymmetric D-Brane Dynamics,'' Phys. Lett. B417 (1998)
233, hep-th/9709063; ``Scale Independent Spin Effects in D-Brane
Dynamics,'' hep-th/9801183.}

\lref\barrio{M.~Barrio, R.~Helling and G.~Polhemus, 
``Spin-Spin Interaction in Matrix Theory,'' J. High Energy Phys. 05
(1998) 012, hep-th/9801189.}

\lref\myers{R.~C.~Myers and M.~J.~Perry,``Black Holes in Higher Dimensional Spacetimes'', 
Ann. Phys. 172 (1986) 304.}

\lref\tod{K.~P.~Tod, ``All Metrics Admitting Supercovariantly Constant Spinors,''
Phys. Lett. 121B (1983) 241.}

%%%%%%The End of References%%%%%%%%%

%%%%%%%%%% Title Page %%%%%%%

\Title{\vbox{\baselineskip12pt
\hbox{UMHEP-455, HUTP-98/A071}
\hbox{hep-th/9811037}}}
{\vbox{\centerline{\titlerm The Spin of the M2-Brane and}
\medskip
\centerline{\titlerm Spin-Spin Interactions via Probe Techniques}
 }}
\centerline{Vijay~Balasubramanian${}^a$\foot{vijayb@pauli.harvard.edu},
David~Kastor${}^b$\foot{kastor, traschen@phast.umass.edu}, 
Jennie~Traschen${}^{b2}$  
and K.~Z.~Win${}^b$\foot{kzwin@hotmail.com}}
\bigskip
\centerline{\it ${}^a$Lyman Laboratory of Physics, Harvard University, 
Cambridge, MA 02138 USA}
%\medskip
\centerline{\it ${}^b$Department of Physics and Astronomy, 
University of Massachusetts, Amherst, MA 01003 USA}
%\medskip
\medskip
\centerline{\bf Abstract}
\medskip
The 256 dimensional M2-brane multiplet contains solitons of many
different intrinsic spins. Using the broken supersymmetry
transformations of the M2-brane, we find supergravity solutions which
explicitly display these spins.  This amounts to quantizing the
fermionic zero modes and computing the back reaction on the metric and
gauge potential.  These spacetime fields are therefore operator valued
and acquire a conventional classical meaning only after taking
expectations in given BPS states.  Our spinning spacetimes are not
of the standard Kerr form -- there is a non-vanishing gravitino.
Nevertheless, the solutions have angular momentum and magnetic dipole
moments with a g-factor of $2$. We use probe techniques to study
scattering of spinning BPS M2-branes. The static interactions cancel
between like-sign branes at leading order, but there are static
spin-spin forces between branes and anti-branes.   The general
probe-background Lagrangian contains gravitational spin-spin and
magnetic dipole-dipole forces, as well as gravitino exchanges which
allow branes to change fermion number.
%\draft
\Date{November, 1998}
\vfill\eject
%%%%%%%%%%%%%%%

\newsec{Introduction}  

\noindent
M$2$-branes and M$5$-branes, the basic solitonic excitations of
M-theory, are each invariant under half of the theory's $32$
supersymmetries.  The action of the remaining $16$ broken
supersymmetries yields, in each case, a BPS multiplet of
$2^8=256$ states, half of which are fermionic and half bosonic.  In
analogy with the spin states of a massive particle, the BPS states of
spinning M-branes fall into representations of a ``little group" of
spatial rotations which leave invariant the world-volume of the brane.
For the M$2$-brane the little group is $SO(8)$, while for the
M$5$-brane it is $SO(5)$.

There are a number of questions we can ask about these BPS states.
For example, at the level of semi-classical spacetime solutions, we
would like to know how the spin of the brane is reflected in the
gauge, gravitational and gravitino fields.  We might expect that,
since the spins are quantum mechanical, of order $\hbar$, the
classical geometry would simply be insensitive to the spin state.
Another possibility is that the different BPS states are described by
Kerr-type spinning M-branes as in \youm , with appropriately quantized
values of the angular momentum.  A third possibility, however, turns
out to be correct.

One can actually generate the spacetime fields of an arbitrary BPS
state by acting iteratively with the broken generators
on the purely bosonic M-brane solution.  Schematically, if $\Phi$
denotes the spacetime fields of the bosonic solution and
$\delta_\epsilon$ the action of the broken supersymmetry generators
with spinor parameter $\epsilon$, the higher spin
states are given by
\eqn\newsoln{
\Phi^\prime = e^{\delta_\epsilon}\Phi=\Phi+\delta_\epsilon\Phi+\half
\delta_\epsilon\delta_\epsilon\Phi + \cdots}
This technique was first used by Aichelberg and Embacher to study the
BPS multiplet of states based on the extreme Reissner-Nordstrom black
holes of $D=4$, $N=2$ supergravity \aezero .  More recently, it has
been applied by Duff et. al.  to compare the dipole moments of black
holes and string states in $D=4$, $N=4$ string compactifications
\duffone\ and to calculate the g-factor for the D$0$-branes of $D=10$,
type $IIA$ supergravity \dufftwo .  In Sec. 2.1, we will carry out the
expansion \newsoln, starting from the bosonic M2-brane, to second
order in the fermionic parameter $\epsilon$.  Following \aezero , we
call this the {\it superpartner solution}.  The superpartner solution
allows us, among other things, to determine the analogue of the
gyromagnetic ratio for states in the M$2$-brane BPS supermultiplet.

The spinor parameter $\epsilon$ in the expansion \newsoln\ is naively
a Grassmann quantity.  This has the desireable effect that the series
\newsoln\ truncates at a finite order. However, it also leads to an
interpretational problem, not addressed in \aezero\ and \dufftwo ,
which needs clarification.  What does it mean for the spacetime fields
to depend on a Grassmann valued parameter?  What values do the fields
actually take at a given spacetime point?

In order to resolve this issue, it is helpful to think about an
analogous bosonic construction first.  Any localized soliton has
bosonic zero-modes, which correspond to broken translation
symmetries. The parameters, called collective coordinates or moduli,
associated with these zero-modes simply specify the center of mass
position of the soliton. Often, as in the case of M-branes written in
multi-center form, we know the solution exactly as a function of these
bosonic parameters. However, if we did not, starting from a reference
solution we could generate it via an expansion of the form
\newsoln, using the generators of broken translations.

The fermionic case should be thought of in the same way. The $16$
broken supersymmetries correspond to $16$ fermionic zero-modes.  The
expansion \newsoln\ gives an exact expression for the soliton as a
function of the fermionic parameters associated with these modes.
However, there is a crucial difference between bosonic zero-mode
parameters and fermionic ones, which was first discussed in \jackiw\
\foot{See also \harvey\ for a good 
discussion in the context of monopoles.}. Fermionic zero-mode
parameters satisfy non-trivial anti-commutation relations and must be
realized as operators acting on a space of quantum
states\foot{Bosonic moduli sometimes also require quantization before
they make sense.  For example, the dyon rotor of BPS monopoles is a
collective coordinate in a $U(1)$ subgroup of the gauge group \harvey\
.  We can imagine endowing the monopole with momentum along the $U(1)$
circle.  Classically any momentum is allowed, but we know that quantum
mechanics dictates integral momenta.  This is, of course, necessary
since the momentum translates into the electric charge of the
monopole.}.  In the case of fermionic zero-modes arising from broken
supersymmetry, this space is the BPS multiplet of spin states.

In the present case, the gravitino field $\psi_m$ has an expansion
\eqn\gravitino{
\psi_m=a_i\psi^i_{m,0} + \hbox{ non-zero modes},\quad i =1,\dots, 16.
}
where $\psi^i_{m,0}$ are zero-modes - linearized solutions to the
field equations in the background of the bosonic M2-brane.  Upon
quantization of the gravitino field, the possible soliton states form
a representation of the algebra of zero mode operators
$\{a_i,a_j\}=\delta_{ij}$, which follows from the canonical
anti-commutation relations of the gravitino field.  Acting with the
broken supersymmetries on the bosonic M2-brane produces the zero modes
of $\psi_m$.  So, up to an important normalization factor, which we will
discuss below, the 16 non-zero components of $\epsilon$, the spinor
parameter of broken supersymmetry, should be identified with the
coefficients $a_i$ and satisfy the same algebra.

Rather than being Grassmann valued, the metric and other spacetime
fields are now seen to be operator valued.  The operators act on the
$256$ dimensional space of BPS states.  In order to obtain c-number
values for the spacetime fields at a given point, we take an
expectation value in a specific BPS state.  This partial quantization
of the spacetime fields comes about because we have effectively
quantized the zero-mode sector of the gravitino, and are finding the
back-reaction on the metric and other spacetime fields.  We are not
referring here to quantum gravity corrections embodied in, say, higher
curvature corrections to the Einstein action.  Merely, quantizing the
zero-modes inevitably leads to back-reaction effects that can modify
the long-range fields even in regions of small curvature.

It may be the case that the operator valued corrections to the bosonic
fields in \newsoln\ do not vanish in {\it any} of the BPS states. In
this case, the original bosonic solution is not exact in any quantum
state - turning on quantum mechanics inevitably produces corrections
to the long-distance fields of the solution.  This is the case for the
D$0$-brane found in \dufftwo\ -- there is no such thing as a spinless
D0-brane in ten dimensions and the semiclassical solutions display
this feature.  In Sec. $2.2$ , we give a construction of the BPS
multiplet of states for the M2-brane. This allows us both to determine
the $SO(8)$ spin content of the multiplet and to evaluate the
spacetime fields in any given state.  Unlike the D0-brane, the
M2-brane does have a singlet state under the transverse rotation
group.  So the original bosonic solution is still valid for a particular
choice of state for the fermionic moduli.

A second set of questions, which we will address, involves the
scattering of spinning M2-branes.  Again following techniques
developed by Aichelberg and Embacher in the series of papers
\refs{\aezero,\aeone,\aetwo,\aethree,\aefour}, in Sec. 3 we study interactions
of spinning M2-branes by examining the effective action for a probe in
the superpartner spacetime backgrounds found in Sec. 2.  The
interaction lagrangian has a purely bosonic piece -- the potential
energy between two M2-branes. This vanishes for like-sign, parallel,
static branes, in accord with BPS force cancellation.  We study the
remaining spin dependent interactions at leading order in a large
separation limit. One of the new terms describes the exchange of a
gravitino between the two branes, which changes the fermion number and
spin of both the background and the probe.  There are also gauge
dipole-dipole interactions and a gravitational spin-spin interaction.
All of these exchanges either cancel, or vanish identically, if the
probe and background have the same sign charge, thereby maintaining
BPS force balance.

\newsec{Spinning M$2$-Branes}
\noindent
First we summarize the formalism.  Further details can be found in
\dewit\ whose conventions and notation we follow.  The metric is of
mostly positive signature, and Dirac matrices are defined by
$\big\{\gamma^{\hat m},\gamma^{\hat n}\big\}=2\eta^{\hat m\hat n}$ and
$\gamma_m=e_m^{\ \hat n}\gamma_{\hat n}$.  Hats indicate orthonormal
frame indices, and $e^{\ \hat{n}}_m$ is the vielbein.
Letters from the middle of the Roman alphabet $\{m,n,p,q,r,s\}$ will
index all 11 coordinates, while letters from the beginning of the
alphabet $\{a,b,c,d\}$ index the directions $\{x^0,x^+,x^-\}$ parallel
to the M2-brane.  Greek letters indicate the directions
$\{x^1,\cdots,x^8\}$ transverse to the brane, and
$\gamma^{mn\dots}$ indicates a product of gamma matrices with all
indices different. Square brackets and round brackets will
indicate (anti)-symmetrization with unit weight\foot{For example,
$A_{[mn]} = (1/2!)(A_{mn} - A_{nm})$}.  With these definitions and the
convention $\gamma^{\hat 0\hat +\hat -\hat 1\hat 2\ldots \hat 8}=1$, 
the eleven dimensional
supersymmetry transformations are:
\eqnn\susyA
\eqnn\susyB
$$\eqalignno{ \delta A_{mnp}=&-6\bar\epsilon\, \gamma_{[mn}\psi_{p]},
%,\hskip4mm
~~~;~~~
\delta e_m^{\ \hat n}=2\bar\epsilon\, \gamma^{\hat n}\psi_m &\susyA\cr
\delta\psi_m=&\left[\partial_m-
\frac 14 \omega_m^{\ \hat n\hat p}\gamma_{\hat n\hat p}+
T_m{}^{npqr}F_{npqr}\right]
\epsilon &\susyB\cr}
$$
where 
$T_m{}^{npqr}\equiv
\frac 1 {288} \left(\Gamma_m{}^{npqr}-8\delta_m^n\Gamma^{pqr}\right)$
and $\epsilon$ is an anticommuting Majorana spinor.  We make frequent
use of the property
\eqn\majorana
{
\bar{\chi} \, \gamma^{\hat{n}_1 \cdots \hat{n}_k} \, \phi
= (-1)^{k(k+1)/2}
\bar{\phi} \, \gamma^{\hat{n}_1 \cdots \hat{n}_k} \, \chi
}
where $\chi$ and $\phi$ are Majorana spinors.  This implies in
particular that $\bar{\chi} \, \gamma^{\hat{n}_1 \cdots \hat{n}_k} \,
\chi = 0$ for $k=1,2,5,6,9,10$.  The supercovariant spin connection is
the (algebraic) solution of the supertorsionless equation
%%%%%%
\eqn\torsionless
{
de^{\hat{m}} - \omega^{\hat{m}}_{~~{\hat{n}}}
\wedge e^{\hat{n}} = \bar{\psi} \wedge \gamma^{\hat m} \psi
}
%%%%
Defining $O^{\hat{m}\hat{n}\hat{p}} = 
e^{\hat{m}q} e^{\hat{n}r} [ \partial_{[q} e^{\hat{p}}{}_{r]} -
\bar{\psi}_{[q} \gamma^{\hat p} \psi_{r]}]$ we find the spin connection:
\eqn\spinconnection{
\omega^{\hat{m}\hat{n}\hat{p}} = O^{\hat{m}\hat{p}\hat{n}}
+ O^{\hat{n}\hat{p}\hat{m}} + O^{\hat{n}\hat{m}\hat{p}}
}
Finally, the supercovariant gauge field strength is
\eqn\fieldstrength
{
F_{mnpq}=4\partial_{[m}A_{npq]}+12\bar\psi_{[m}\gamma_{np}\psi_{q]}\> .
}

\subsec{The Superpartner Solution}
\noindent
The membrane solution  of Duff and Stelle is described by the following
fields:
\eqn\membrane{ 
ds^2={\eta_{ab}\over f^2}dx^adx^b+f\delta_{\alpha\beta}
dx^{\alpha}
dx^{\beta}
%,\hskip3mm 
~~~;~~~
A_{abc}=
-{s\over f^3}\varepsilon_{abc}
%,\hskip2mm 
~~~;~~~
s=\pm 1
%,\hskip2mm 
~~~;~~~
\psi_m=0
}
where $\varepsilon$ is totally antisymmetric in its indices with
$\varepsilon_{0+-}=-\varepsilon^{0+-}=1$ and $s$ is the sign of the
electric charge.  The metric function $f^3$ is a harmonic function of
the transverse coordinates,
$\delta^{\alpha\beta}\partial_\alpha\partial_\beta f^3=0$. The
asymptotically flat solution is
\eqn\fdef
{
f=\left(1+\sum_i {3M_i\over |\vec r-\vec r_i|^6}\right)^{{1 \over 3}}
}
where $\vec r=(x^1,\dots,x^8)$.  If the positions $\vec r_i$'s are
sufficiently well separated that there is a local asymptotically flat
region around each, then $M_i$ has the interpretation of the mass of
each individual membrane and $Q_i\equiv sM_i$ is the electric charge.
In any case $M\equiv\sum M_i$ can be interpreted as the total mass of
the spacetime and $Q\equiv sM$ is the total electric charge.  Defining
$e^{\hat{m}}= e^{\hat{m}}_{\ n} dx^n$, the non-zero components of
vielbein are
\eqn\viela
{
e^{\hat{a}} = f^{-1} \delta^{\hat a}_b dx^b ~~~~~~;~~~~~~ 
e^{\hat{\alpha}} = f^{1/2}\delta^{\hat\alpha}_\beta dx^{\beta}. 
}
The connection one-form $\omega^{\hat{n}\hat{p} } =
\omega_{m}^{~~\hat{n}\hat{p}} dx^m$ has nonvanishing components
\eqn\conna
{
\omega^{\hat{a}\hat{\alpha}} = f^{-5/2} (\partial_{\beta} f) 
\delta^{\hat a}_b\delta^{\hat\alpha}_\beta dx^b
~~~~~~;~~~~~~
\omega^{\hat{\alpha}\hat{\beta}} = {1\over 2} f^{-1/2}
(\partial_\eta f) dx^\rho \left[ \delta^{\hat\alpha}_\rho
\delta^{\hat\beta\eta} - \delta^{\hat\alpha\eta}\delta^{\hat\beta}_\rho
\right].
}
We will start from this bosonic solution to generate the spinning
superpartners\foot{The coordinate indices on $\delta^{\hat a}_b$ and
$\delta^{\hat\alpha}_\beta$ are raised and lowered using the flat
Minkowski metric.}.

\mybullet{Supersymmetry} 
The multimembrane solution preserves one half
of the supersymmetry in the sense that $\delta\psi_m=0$ if
$\epsilon=f^{-\half}\lambda$, with constant Majorana spinor $\lambda$
satisfying $(1+s\tilde\gamma)\lambda=0$.   Here
$\tilde\gamma\equiv\gamma^{\hat 0\hat +\hat -}$ with ${\tilde
\gamma}^2=1$ and $\Tr\tilde\gamma=0$, setting half of the independent
components of $\lambda$ to zero.  The superpartner of the membrane is
obtained by considering supersymmetry transformations with $\lambda$
chosen so that
%%%%%%
\eqn\fzero
{
(1- s\tilde\gamma)\lambda=0
}
%%%%%%%%%
 which implies that $\delta\psi_m\ne 0$.
To generate superpartners one starts with the membrane background
\membrane\ and applies the broken supersymmetry transformations
 to find a nonzero gravitino field.  Then the gravitino field is
inserted into \susyA\ to obtain the corrections to other fields.  This
iterative process can be repeated to higher order in $\lambda$ and
will terminate after a finite number of terms because $\lambda$
anticommutes.  However, we will content ourselves with the order
$\lambda^2$ terms.

\mybullet{Superpartners} 
Since there is a local supersymmetry,
the spinor $\epsilon$ parametrizing the broken generators can be
related to $\lambda$ by an arbitrary multiplicative function
\hbox{$\ce=\ce(x^\alpha)$} which goes to one at infinity.
%Because $\delta\psi_m\ne 0$, $\epsilon$ needs be the same as $\lambda$
%only up to a multiplicative factor \hbox{$\ce=\ce(x^\alpha)$}.  This
%arbitrariness is nothing but local supersymmetry at play.  
Choosing $\ce$ is the same as fixing a gauge for local supersymmetry
transformations.  One family of gauge choices is $\epsilon
=f^{-\delta} \lambda$.  One can check that the broken supersymmetry
variation of the gravitino field is then normalizable for any value of
$\delta >0$. One can also check explicitly that the supercharge, given
by a surface integral at infinity \teit , is independent of the value
of $\delta$, as it must be. In parallel with the unbroken generators,
we will take $\epsilon =f^{-1/2}\lambda $ below. The first order
variation gives the gravitino:
\eqn\gravitino
{
\psi = -f^{-3}(\delta^{\hat\alpha\beta} \, \partial_\beta f) \, 
\gamma_{\hat b\hat\alpha}\, \lambda \, \delta^{\hat b}_a \, dx^a
+f^{-3/2}\left[ -\partial_\alpha f+\half\delta_\alpha^{\hat\rho} \, 
\delta^{\beta\hat\sigma}\, \partial_\beta
f\gamma_{\hat\rho\hat\sigma}\right]\, \lambda \, dx^\alpha. } 
%
%where $f_\alpha=f,_\alpha$.  
Iterating the transformation 
yields the vielbein to order $\lambda^2$:
\eqn\vielbein{\eqalign{
e^{\hat a}&=f^{-1} \, \delta^{\hat a}_b \, dx^b+\half
f^{-2}(\delta^{\hat\rho\beta} \, 
\partial_\beta f)\left(\bar\lambda \, 
\gamma^{\hat a}_{\ \hat\alpha\hat\rho} \,
\lambda\right)\delta^{\hat\alpha}_\sigma \, dx^\sigma\cr 
e^{\hat\alpha}&=f^{\half} \, \delta^{\hat\alpha}_\beta \, dx^\beta -
f^{-7\over 2}(\delta^{\hat\rho\beta} \, \partial_\beta f)
\left(\bar\lambda \, \gamma^{\hat\alpha}_{\  \hat b\hat\rho} \,
\lambda\right) 
\delta^{\hat b}_a \, dx^a.\cr
}}
To this order in $\lambda$, only off-diagonal components with one
index tangent to the brane and one transverse, receive corrections
\eqn\metric{
g_{a\alpha}={3\over 2}f^{-3}\ \delta_a^{\hat b} \,
\delta_\alpha^{\hat\beta}\  
(\delta^{\hat\rho\eta} \, \partial_\eta f)\ 
\left(\bar\lambda \, \gamma_{\hat b\hat\beta\hat\rho} \,
\lambda\right), 
}
%{g_{c\alpha }={3\over 2 f^3}f_\beta\hat{\Lambda}_{c\alpha }{}^\beta}
%
while the gauge field has corrections to the following components
\eqn\gaugepert{\eqalign{
A_{ab\alpha}&={-3s\over 2} \, f^{-4}\ \varepsilon_{ab\hat c}\ 
\delta^{\hat\rho\alpha} \, \delta^{\hat\sigma\beta} \, \partial_\beta  
f\ \left(\bar\lambda \, \gamma_{\hat\rho\hat\sigma}^{\ \ \hat c} \,
\lambda\right )\cr 
A_{\alpha\beta\rho}&=-{3\over 2} \, f^{-1}\ 
\delta_\alpha^{\hat\mu}\,\delta_\beta^{\hat\nu}\,\delta_\rho^{\hat\eta}
\,\delta^{\hat\sigma\chi}
\ \partial_\chi f\left(\bar\lambda \,
\gamma_{\hat\mu\hat\nu\hat\eta\hat\sigma}  \, \lambda\right).\cr
}}
As a check, note that condition $|Q|=M$ continues to hold as
expected for members of the BPS supermultiplet.

\mybullet{Angular Momentum and Dipole Moments}
For membrane spacetimes, the long-distance limit of the off-diagonal
metric components $g_{a\alpha}$ determine an angular momentum current
$J_a^{\ \alpha\beta}$, which carries a world-volume vector index in
addition to the usual pair of transverse indices specifying a plane of
rotation.   This is given by \myers
\eqn\spin{
g_{a\alpha}{\buildrel {r\to\infty}\over \longrightarrow}\, -{8\pi
J_{a\alpha\beta}x^\beta \over \Omega_7 r^8},} 
where $\Omega_7$ is the
area of a unit $7$-sphere.  From the world-volume perspective of the
brane, the angular momentum current arises because the membrane
effective Lagrangian is invariant under transverse rotations. There
is a corresponding conserved angular momentum current with the index
structure of $J_a^{\
\,\alpha\beta}$, in analogy with the angular momentum of a particle
$J_0^{\ \,\alpha\beta}$.  This current is registered in the
long-distance metric as in \spin.  For the superpartner spacetimes,
the angular momentum current coming from the long distance limit of
\metric\ is:\foot {Because gamma matrices with
frame and coordinate indices are identical in the long distance limit,
we have dropped hats from the gamma matrices in this and other
expressions below when appropriate.}
\eqn\genang{
J_a^{\ \,\alpha\beta}={9\Omega_7 M\over 8\pi}
\left(\bar\lambda \gamma_{a}^{\ \ \alpha\beta}\lambda\right)
}
which is a bilinear in the spinor $\lambda$.  We will see in Sec. 2.2
that $J_0^{\alpha\beta}$ generates rotations in the space of fermionic
zero-mode states.  The full angular momentum current determines the
gravitational spin-spin interaction between M2-branes, as shown in
Sec. 3.

Two different dipole moment tensors can be extracted from the long
distance limit of the gauge field components in \gaugepert . The
long-distance limit of $A_{ab\alpha}$ yields a dipole moment tensor
$\mu_a^{\ \,\alpha\beta}$, having the same index structure as the
angular momentum current:
\eqn\firstdipole{A_{ab\alpha}{\buildrel
r\to\infty\over\longrightarrow}\, 
{8\pi\varepsilon_{ab}^{\ \ \ c}\mu_{c\alpha\beta}x^\beta\over \Omega_7 r^8}.}
From \gaugepert\ we find
\eqn\firstvalue{
\mu_a^{\ \,\alpha\beta}={9s\Omega_7 M\over 8\pi}
\left(\bar\lambda \gamma_{a}^{\ \ \alpha\beta}\lambda\right).
}
If we define a g-factor via the relation $\mu_a^{\
\,\alpha\beta}=(gQ/2M)J_a^{\ \,\alpha\beta}$, using  $Q=sM$ we find
that $g=2$ for the superpartner spacetimes.
From the long distance limit of $A_{\alpha\beta\rho}$, we can define a
dipole moment tensor $\mu_{\alpha\beta\rho\sigma}$ with four
transverse indices:
\eqn\seconddipole{
A_{\alpha\beta\rho}{\buildrel r\to\infty\over\longrightarrow}\, 
{8\pi \mu_{\alpha\beta\rho\sigma}x^\sigma\over \Omega_7 r^8}.
}
In the superpartner spacetimes this has the value:
\eqn\secondvalue{
\mu_{\alpha\beta\rho\sigma}={9\Omega_7 M\over 8\pi}
\left(\bar\lambda\gamma_{\alpha\beta\rho\sigma}\lambda\right).
}
We will see in Sec. 3 that both the dipole moments $\mu_a^{\
\,\alpha\beta}$ and $\mu_{\alpha\beta\rho\sigma}$ mediate
interactions between branes.

\mybullet{Further Discussion of Superpartner Spacetimes}
There is an interesting issue which arises if the starting point for
the superpartner construction is taken to be a multi-M2-brane
spacetime, rather than just a single brane.  Acting with the broken
supersymmetry generators on the multi-brane spacetime produces only a
single overall BPS multiplet of spin states, rather than a
multiplet for each object.  This makes sense, if we think about the
analogous bosonic case, acting with the broken translation symmetries
on a multi-brane spacetime. All the branes are translated together and
so we cannot access the moduli which vary the relative positions of
the branes via the broken translation symmetries. Similarly, we cannot
access fermionic zero-modes for each brane (if they exist) using the
broken supersymmetries. In the bosonic case, of course, we know the
multi-center solutions which allow us to vary the relative positions
of branes.  It is unclear whether, or not, in the case of fermionic
zero-modes, independent spins may be associated with each brane.

There are pieces of evidence which point in both directions on this
question.  At the classical level in  four dimensions, there are the
Israel-Wilson-Perjes \iwp\ spacetimes which are supersymmetric \tod\
and describe multiple objects with arbitrary individual angular
momenta. In \bkt\ it is shown that these IWP objects satisfy a
balance between gauge and gravitational spin-spin forces, similar to
our result below in Sec. 3.  On the other hand, in M(atrix) model and
other calculations of spin-spin forces between D0-branes
\refs{\harveyspin,\krausspin,\morales,\barrio}, nonzero forces have
been found between static objects, indicating that an exact force
balance occurs only for certain combinations of the individual spins.
Indeed, we expect that exact force cancellation occurs only if the
spins of the individual objects assemble into an overall BPS state and
so static solutions should not permit arbitrary individual spins for
multi-center M2-branes.

\subsec{Spin Content of the BPS Multiplet}

\noindent
The metric and other fields of the superpartner spacetimes depend on
the spinor parameter $\lambda$, which satisfies the projection
$(1-s\tilde\gamma)\lambda=0$.  As we discussed in the introduction,
the non-zero components of $\lambda$ satisfy non-trivial
anti-commutation relations and must therefore be realized as operators
acting in a quantum mechanical space of states, which are the 256
different spin states of the BPS multiplet. In this section we will
make these observations more concrete and give an explicit
construction of the space of states.

We begin by fixing a representation of the $SO(10,1)$ Dirac matrices,
which is adapted to the decomposition of $SO(10,1)$ representations
into representations of
the $SO(2,1)\otimes SO(8)$ subgroup of Lorentz transformations in the
tangent directions and rotations in the transverse directions to the
brane. We take\foot{We are also omitting hats from indices in this
subsection.}
\eqn\diracrepn{\eqalign{&\gamma^0=i\sigma^2\otimes {\rm I_{16}},\qquad
\gamma^+=\sigma^1\otimes {\rm I_{16}},\qquad \gamma^-=\sigma^3\otimes \bar{\Gamma},\cr
&\gamma^\alpha=\sigma^3\otimes \Gamma^\alpha,}}
where $\sigma^k$ are the Pauli matrices, $I_{16}$ is the $16$
dimensional identity matrix and $\Gamma^\alpha$ are $SO(8)$ Dirac
matrices with $\bar\Gamma=\Gamma^{12345678}$. To make things entirely
explicit, for the $\Gamma^\alpha$ we take the representation
\eqn\moredirac{\eqalign{&\Gamma^\alpha=\left(\matrix{0&\eta^\alpha\cr
\xi^\alpha & 0}\right ),\qquad \xi^\alpha=(\eta^\alpha){}^t, \cr
&\eta^1=\varepsilon\otimes\varepsilon\otimes\varepsilon ,
\qquad \eta^2=1\otimes\sigma^1\otimes\varepsilon,\qquad \eta^3=1\otimes\sigma^3\otimes\varepsilon ,\cr &
\eta^4=\sigma^1\otimes\varepsilon\otimes 1,\qquad \eta^5=\sigma^3\otimes\varepsilon\otimes 1,
\qquad \eta^6=\varepsilon\otimes 1\otimes\sigma^1,\cr
&\eta^7=\varepsilon\otimes 1\otimes\sigma^3,\qquad \eta^8=-1\otimes
1\otimes 1,\cr}}
with $1=I_2$ and $\varepsilon= i\sigma^2$, which gives 
\eqn\gammabar{\bar{\Gamma}=
\left (\matrix{{\rm I}_8 & 0\cr 0 & -{\rm I}_8} \right ).}
We then have $\tilde{\gamma}={\rm I}_2\otimes \bar{\Gamma}$. 
For $s=1$, solutions to the projection condition $\tilde\gamma\lambda=s\lambda$ can then 
be written in terms of a pair of positive
chirality $SO(8)$ spinors $\rho$ and $\chi$ as
\eqn\spinors{\lambda=\left(\matrix{\rho\cr 0\cr\chi\cr 0}\right).}
The zero mode part of the canonical anti-commutation relations for
the gravitino field determines an algebra for the $16$ components
of $\rho^A$ and $\chi^A$ of the form
\eqn\relations{\{\rho^A,\rho^B\}=\{\chi^A,\chi^B\}=N^2\delta^{AB},
\qquad \{\rho^A,\chi^B\}=0,}
with $N^2$ a normalization factor to be determined below.
Alternatively, since the gravitino field is linear in $\lambda$, this
same algebra for $\rho^A$ and $\chi^B$ arises from the
anti-commutation relation of broken supercharges, given in terms of
surface integrals evaluated at infinity \teit\ in the superpartner
spacetimes \aezero .
%As in \aezero\ the supercharge, expressed in terms of the long distance limit of 
%the gravitino field \teit , is
%proportional to $\lambda$.
%The supercharge of the spacetime which is generated by the gravitino
%\gravitino  , is proportional to the spinor $\lambda$ \teit  ,\aezero  .
%Therefore the sixteen independent choices of $\lambda$ will satisfy
%particular anticommutation relations, which follow from the SUSY algebra.

These anticommutation relations imply that the $16$ operators $\rho^A
,\ \sigma^B$ may be represented by $2^{16/2}=256$ dimensional $SO(16)$
Dirac matrices. The $256$ dimensional space of states on which these
act is the BPS multiplet of spin states for the superpartner
spacetimes.  An explicit constructions of these operators is obtained
by working with the $8$ complex combinations
$\Lambda^A=(\rho^A+i\sigma^A)/\sqrt{2}N$, which together with their
hermitian conjugates $\Lambda^{A\dagger}$ satisfy a fermionic
creation/annihilation algebra,
\eqn\newrelns{\{\Lambda^A,\Lambda^{B\dagger}\}=\delta^{AB},
\qquad \{\Lambda^A,\Lambda^{B}\}=\{\Lambda^{A\dagger},\Lambda^{B\dagger}\}=0.}
$\Lambda^A,\Lambda^{B\dagger}$ then act straightforwardly on the
$256$ dimensional space of states constructed from the vacuum
$|0>$, with $\Lambda^{A}|0>=0$. The vacuum state $|0>$ is an $SO(8)$
singlet\foot{This is in contrast to the D0-brane which has no singlet
states under $SO(9)$, its transverse rotation group \dufftwo .}.  
 The first level of excited states $\Lambda^{a\dagger}|0>=0$
form an $SO(8)$ spinor with chirality $s$, and so forth for higher
spin states.   
The original $SO(10,1)$ spinor $\lambda$, expressed in terms of $\Lambda$ and $\Lambda^\dagger$,
is then given by
\eqn\spinors{\lambda=
{1\over\sqrt{2}N}\left(
\matrix{\Lambda +\Lambda^\dagger\cr 0\cr -i(\Lambda-\Lambda^\dagger)\cr 0}
\right).}
We can then express the fermion bilinears 
$\bar\lambda\gamma_{\hat m\hat n\dots}\lambda$ which appear in the metric, gauge and gravitino
fields of the superpartner spacetime
in terms of the $8$ creation and annihilation operators 
$\Lambda^{A\dagger},\ \Lambda^A$ which act on the BPS multiplet of
states. 

In order to determine the value of the normalization constant $N$
appearing in \spinors , we impose the physical requirement that the
spacetime angular momentum $J_0^{\ \alpha\beta}$ in eq. \genang\ act
as the generator of $SO(8)$ rotations on the BPS space of states,
which it has the correct form to be.  This simply requires that the
angular momentum recorded in the long distance behavior of the metric
is precisely that carried by the matter fields, in this case the
angular momentum of the fermion zero-modes.  Expressing $J_0^{\
\alpha\beta}$ in terms of creation and annihilation operators, we get
\eqn\angmom{ J_0^{\ \alpha\beta}= {18\Omega_7 M\over\pi \hbar N^2}\left({\hbar\over 4}
\Sigma_{+AB}^{\alpha\beta}\Lambda^{A\dagger}\Lambda^B\right),}
where
$\Sigma_+^{\alpha\beta}=\half(\eta^\alpha\xi^\beta-\eta^\beta\xi^\alpha)$.
The operator within the parentheses in \angmom\ is the properly
normalized $SO(8)$ rotation generator.  The normalization constant $N$
must then be $N^2=18\Omega_7 M/\pi\hbar$.  This implies that when
corrections to the spacetime fields are expressed in terms of properly
normalized creation and annihilation operators, the factors of $M$ and
$Q$ disappear.  Rather, the angular momentum corrections are
proportional to $\hbar$, as we should expect for the quantized
zero-mode system.  Indeed, upon taking expectation values in different
zero-mode states, \angmom\ correctly registers the expected intrinsic
spin in the long-range, classical gravitational fields.  The dipole
moment $\mu_{\alpha\beta\rho\sigma}$ may also be expressed in terms of
the creation and annihilation operators.

The reader may worry that an order $\hbar$ correction to the metric of
this kind cannot be consistently maintained when there are order
$\hbar$ higher curvature corrections to the equations of motion.
However, this is not a concern because the metric is becoming flat at
large distances and the higher curvature corrections will be
infinitesimal.

\newsec{Spin-Spin Interactions}
\noindent
In this section we will study the spin dependent interactions between
a pair of M$2$-branes by treating one of the branes as a probe
propagating in the spacetime fields of the other. The supersymmetric
world volume action for the probe M$2$-brane \maction\ is
\eqn\action{
S[Z(\zeta)]=\int\left[ -\sqrt{-\det G(Z)}-\frac 16s'\varepsilon^{abc}
\Pi_a^A\Pi_b^B\Pi_c^CB_{CBA}(Z)\right] d^3\zeta
=\int L\,d^3\zeta \> ,
}
where we have inserted $s'=\pm 1$, to account for the sign of the the
probe brane charge, in front of the pull-back of the super gauge field
$B_{MNP}$.  The sign of this second term can be checked by requiring
that in the bosonic limit($\lambda,\theta\to 0$) the static force
should vanish for like-charge ($s'=s$) probes and backgrounds. Here
$\zeta^b$ are the three world volume coordinates of the probe brane
while $Z^M=(X^m,\theta)$ are the superspace coordinates of the probe
and are functions of $\zeta^b$.  The indices $A,B$ denote superspace
frame indices and $M,N$ are superspace coordinate indices.  The
quantity $\Pi_c^A=(\partial Z^M/\partial\zeta^c)E_M^A$ is the
pull-back of the supervielbein to the membrane world volume, and
$G_{ab}=\Pi_a^{\hat m}
\Pi_b^{\hat n}\eta_{\hat m\hat n}$  the induced metric on the probe.

Our calculation requires explicit expressions for the superfields
$B_{MNP}(X^m,\theta)$ and $E_M^A(X^m,\theta)$ in terms of the
component fields $e_m^{\hat n}$, $\psi_m$ and $A_{mnp}$.  Fortunately,
these have recently been given to second order in the fermionic
coordinate $\theta$ in \dewit.  We study the interactions of the probe
with the superpartner backgrounds of Sec. 2.1 in the static limit.
Now there are two sets of fermions, $\lambda$ and $\theta$, in the
calculation, which are realized as operators acting on the spin states
of the background and probe branes respectively.

We will work in static gauge $\partial_aX^b=\delta^b_a$, supplemented by the
$\kappa$-symmetry gauge fixing condition 
%%%%%%%
\eqn\thetafzero
{
(1-s^\prime\tilde\gamma)\theta=0\>,
%\qquad\sigma=\mp 1
}
%%%%%%
which cuts in half the number of fermionic coordinates for the
probe. This gauge condition for the probe is identical to the
condition \fzero\ satisfied by the fermionic parameters of the
superpartner spacetimes. In addition, we will restrict ourselves to
probes at rest and set $\partial_aX^\alpha=\partial_a\theta
=0$. Subject to these conditions, we expand the Lagrangian $L$ in
\action\ out to terms of order $\theta^2\lambda^2$ and accurate to
order in $1/r^8$, where $r$ is the transverse separation between the
background and probe branes. In practice, this will allow us to drop
terms containing $(\partial f)^2$ relative to terms of order
$f\partial^2 f$, simplifying the calculation considerably.

Given this set of conditions, it follows from equation (5.1) of reference \dewit ,
that the following terms may contribute to the induced metric on the probe brane
\eqn\pullbackmetric{\eqalign{
G_{ab}= & \bar g_{ab} + \eta_{\hat c\hat d}\left\{
4\bar e^{\hat c}_{(a}\bar\theta\gamma^{\hat d}\psi_{b)}
-\half \bar e^{\hat c}_{(a}\bar\theta\gamma^{\hat d\hat p\hat q}\theta
\tilde\omega_{b)\hat p\hat q}\right .\cr &\left .
+2 \bar e^{\hat c}_{(a}\bar\theta\gamma^{\hat d}T_{b)}{}^{pqrs}\theta
\tilde F_{pqrs}\right\}.}}
Here, $\bar g_{ab}$ and $\bar e^{\ \hat a}_b$ are the background
metric and vielbein at order $\lambda^0$ given in \membrane\ and
\viela . The quantities $\tilde\omega^{\hat p\hat q}$ and $\tilde
F_{pqrs}$ are the order $\lambda^2$ contributions to the background
spin connection and field strength and are given below.  Some
apparently lower order terms in \pullbackmetric , with the order
$\lambda^0$ spin connection and field strength replacing $\tilde
\omega $ and $\tilde F$, have vanished after making use
of the $\kappa$ projection and Majorana properties satisfied by the
spinor $\theta$.  The $\sqrt{-\det\, G}$ term in the action is then given to
the appropriate order by
\eqn\firstterm{
\sqrt{-\det\, G}=f^{-3} +2f^{-2}\delta_{\hat d}^b\left(\bar\theta\gamma^{\hat d}
\psi_b\right) -{1\over 4}f^{-2}\delta_{\hat d}^b
\left(\bar\theta\gamma^{\hat d\hat p\hat q}\theta\right)
\tilde\omega_{b\hat p\hat q} +
f^{-2}\delta_{\hat d}^b\left(\bar\theta\gamma^{\hat d}T_b{}^{pqrs}
\theta\right)\tilde F_{pqrs}.
}
Here, the purely bosonic $f^{-3}$ term comes  from the determinant of the 
original, order $\lambda^0$ M-$2$-brane metric
\membrane.

We will now see that the second term in the action \action\ 
has a very similar form.
The pull-back of the super gauge field is given in equation (5.2) of reference
\dewit . After plugging in the conditions for static gauge in the
stationary probe, and dropping higher order terms in $1/r$, their
expression reduces to
\eqn\supergaugefield{\eqalign{
-{s^\prime\over 6} \varepsilon^{abc}
\Pi_a^A\Pi_b^B\Pi_c^CB_{CBA}=&{s^\prime\over 6}\varepsilon^{abc}
\left\{A_{abc}-6\left(\bar\theta \, \gamma_{ab} \, \psi_c\right) \right .\cr
& \left .+{3\over 4}\left(\bar\theta \, \gamma_{\hat p\hat q} \,
\gamma_{ab} \, \theta\right) 
\tilde\omega_c^{\ \hat p\hat q} -
3\left(\bar\theta \gamma_{ab}T_c{}^{pqrs}\theta\right)
\tilde F_{pqrs}\right\}.\cr
}}
The 2nd, 3rd and 4th terms above are quite similar to the
corresponding terms in equation \firstterm , with each term in
\supergaugefield\ containing one extra gamma matrix.  Applying the gauge
condition \thetafzero\ brings these to the same form as the
corresponding terms in
\firstterm. Plugging in from \membrane\ for the gauge potential 
$A_{abc}$ and combining \firstterm\ and \supergaugefield\ then gives
for the Lagrangian $L$
\eqn\lagrangian{\eqalign{
L=&(ss^\prime-1)f^{-3} 
-4f^{-2}\delta_{\hat d}^b\left(\bar\theta\gamma^{\hat d}
\psi_b\right) +{1\over 2}f^{-2}\delta_{\hat d}^b
\left(\bar\theta\gamma^{\hat d\hat p\hat q}\theta\right)
\tilde\omega_{b\hat p\hat q}\cr & -2
f^{-2}\delta_{\hat d}^b\left(\bar\theta\gamma^{\hat d}T_b{}^{pqrs}
\theta\right)\tilde F_{pqrs}.
\cr}}
It remains to plug in for the order $\lambda^2$ spin connection and
gauge field strength $\tilde\omega_a^{\hat p\hat q}$ and $\tilde
F_{pqrs}$. Keeping only terms of $O\,(1/r^8 )$, the gravitino squared
terms in \torsionless\ do not contribute. Hence $\tilde \omega $ can
be found from
\eqn\secondorder{\tilde \omega ^{\hat{m}}_{~~{\hat{n}}} \wedge
\bar e^{\hat{n}} =   d\tilde e^{\hat{m}} - \bar\omega^{\hat{m}}_{~~{\hat{n}}}
\wedge \tilde e^{\hat{n}} ,} 
and $\tilde F$ follows from $\tilde A$.
These have non-zero components
\eqn\secondorder{\eqalign{
\tilde\omega^{\hat a\hat\beta}=& -{1\over 4}f^{-5\over 2}
(\delta^{\hat\chi\rho}\partial_\eta\partial_\rho f)
\left(\delta^{\hat\alpha\hat\beta}\delta^\eta_\sigma 
+3\delta^{\hat\alpha}_\sigma\delta^{\hat\beta\eta}\right)
\left(\bar\lambda\gamma^{\hat a}_{\ \hat\alpha\hat\chi}\lambda\right)dx^\sigma\cr
\tilde\omega^{\hat\alpha\hat\beta}=&-{3\over 4}f^{-4}
(\delta^{\hat\chi\rho}\partial_\eta\partial_\rho f)
\left(\delta^{\hat\alpha\hat\mu}\delta^{\hat\beta\eta}-
\delta^{\hat\alpha\eta}\delta^{\hat\beta\hat\mu}\right)
\left(\bar\lambda\gamma_{\hat b\hat\mu\hat\chi}\lambda\right)
\delta_a^{\hat b}dx^a\cr
\tilde F_{ab\alpha\beta}= &{3s\over 2}f^{-4}\varepsilon_{ab\hat c}
\left(\delta_\alpha^{\hat\rho}\delta_\beta^\chi-
\delta_\alpha^\chi\delta_\beta^{\hat\rho}\right)
(\delta^{\hat\sigma\eta}\partial_\chi\partial_\eta f)
\left(\bar\lambda\gamma_{\hat\rho\hat\sigma}^{\ \ \hat c}\lambda\right)\cr
\tilde F_{\alpha\beta\rho\sigma}= & -6f^{-1}(\partial_\chi\partial_{[\alpha} f)
\delta_\beta^{\hat\nu}\delta_\rho^{\hat\psi}\delta_{\sigma]}^{\hat\eta}
\delta^{\hat\mu\chi}
\left( \bar\lambda\gamma_{\hat\nu\hat\psi\hat\eta\hat\mu}\lambda\right).\cr
}} Inserting these in \lagrangian\ and working through gamma matrix
algebra, using the Majorana properties \majorana\ and projection
conditions \fzero\ and \thetafzero\ to simplify, yields our final
result for the Lagrangian in this limit
\eqn\final{\eqalign{
L\equiv & \ L_0+L_1+L_2+L_3\cr
=& \ (ss^\prime-1)f^{-3} +6(ss^\prime-1)f^{-5}\left(\delta^{\hat\alpha\beta}\partial_\beta f\right)
\left(\bar\theta\gamma_{\hat\alpha}\lambda\right)\cr
&+{3\over 4}(ss^\prime-1)f^{-6}\left(\delta^{\hat\psi\rho}\delta_{\hat\beta}^\eta
\partial_\rho\partial_\eta f\right)
\left(\bar\theta\gamma^{\hat\alpha\hat\beta\hat d}\theta\right)
\left(\bar\lambda\gamma_{\hat\alpha\hat\psi\hat d}\lambda\right)\cr
&-{1\over 8}f^{-6}
\left(\delta^{\hat\psi\rho}\delta_{\hat\beta}^\eta
\partial_\rho\partial_\eta f\right)
\left(\bar\theta\gamma^{\hat\alpha\hat\mu\hat\nu\hat\beta}\theta\right)
\left(\bar\lambda\gamma_{\hat\alpha\hat\mu\hat\nu\hat\psi}\lambda\right).\cr
}}
For ease of exposition we have broken the lagrangian up into four
terms - $L_0$, $L_1$, $L_2$ and $L_3$ - each of which we discuss
separately below.  Note that $L_2 $ comes from the second and third
terms in \lagrangian , while $L_3$ comes entirely from the last term.
Of particular interest is the question of whether there is a force
balance between like-charge superpartner states with
$s=s^\prime$. This is obviously the case for all the terms except $L_3$.

\mybullet{$L_0$ -- Bosonic Potential} 
The first term in the Lagrangian \final
\eqn\zero{
L_0=(ss^\prime-1)f^{-3}} 
is zeroth order
in the fermions $\theta$ and $\lambda$.  This is simply (minus) the
potential energy of two purely bosonic, parallel M$2$-branes.  $L_0$
vanishes for $s=s^\prime$ because the gravitational attraction cancels
the charge repulsion -- this is simply the well-known BPS force
balance. In this term, but not the others, the factor $f^{-3}$ is
accurate and the forces balance for all separations, not just to
leading order in a large separation expansion.

The remaining terms in the Lagrangian have been calculated accurate
only to leading order in a long distance expansion.  Hence, we now set
$f=1$ in $L_1$, $L_2$ and $L_3$. In addition the factors
$\partial_\alpha f$ and $\partial_\alpha\partial_\beta f$ should be
understood as being given by their large $r$ limits
\eqn\largerlimit{
\partial_\alpha f=-{6Mx^\alpha\over r^8},\qquad
\partial_\alpha\partial_\beta f= -{6M\delta_{\alpha\beta}\over r^8}
+{48 Mx^\alpha x^\beta\over r^{10}}
}

\mybullet{$L_1$ -- Gravitino Exchange} 
The term $L_1$ is first order in both $\lambda$ and $\theta$ and
is given in the long distance limit by
\eqn\Lone{
L_1=-36(s's-1) {M\over r^7}
\delta^{\hat\beta}_\alpha x^\alpha
\left(\bar\theta\gamma_{\hat\beta}\lambda\right).   
}
This interaction changes the spin states of both the probe and the
superpartner background via gravitino exchange.  Recall that in the
Introduction and Sec. 2.2 we interpreted the parameters $\lambda$ as
operators acting on the fermionic state of the superpartner
spacetime. Operators which are odd in powers of $\lambda$ take bosonic
states to fermionic states and vice-versa.  A similar interpretation
holds for the fermionic parameters $\theta$ of the probe, so that
$L_1$ changes the fermion number of both the probe and the background.

We are dealing with an unusual and seemingly contradictory situation,
in which the probe can alter the state of the background. That this
does not happen to an appreciable extent is the usual definition of
the probe approximation. Here, even though the mass of the background
is macroscopic and much larger than that of the probe, the background
spin is of order $\hbar$ and comparable to that of the probe. We must
therefore treat the spin states of the two objects on the same
footing.  In this case it is necessary to discuss a macroscopic object
with a microscopic spin because if a high mass M2 background is BPS,
its spin is still in the same small $SO(8)$ representations as the BPS
probe, rather than in some very large representation.

\mybullet{$L_2$ -- Gauge and Gravitional Spin-Spin Forces}
Plugging in the long distance limit, the interaction $L_2$ is given by
\eqn\ltwo{
L_2={3\over 4}(ss^\prime-1)\left(
{-6M\delta^{\rho\eta}\over r^8}+{48Mx^\rho x^\eta\over r^{10}}\right)
\delta_\rho^{\hat\chi}\delta_{\hat\beta\eta}
\left(\bar\theta\gamma^{\hat\alpha\hat\beta\hat d}\theta\right)
\left(\bar\lambda\gamma_{\hat\alpha\hat\chi\hat d}\lambda\right).
}
This term clearly shows a force balance for $ss^\prime=1$. However, we
may still inquire as to the nature of the forces which are canceling
each other.  The fermion bilinears in \ltwo\ can be re-expressed in
terms of the angular momentum currents of the probe and background
branes.  So we are seeing a cancellation between gauge and
gravitational spin-spin interactions \wald , similar to the
cancellation occuring \bkt\ in the IWP spacetimes \iwp.

%\vskip 0.2in\noindent {\bf $L_3\ $:}
\mybullet{$L_3$ -- Dipole-Dipole Interactions}
The long distance limit of the term $L_3$ in \final\ is
\eqn\lthree{
L_3= -{1\over 8}\left(
{-6M\delta^{\rho\eta}\over r^8}+{48Mx^\rho x^\eta\over r^{10}}\right)
\delta_\rho^{\hat\psi}\delta_{\hat\beta\eta}
\left(\bar\theta\gamma^{\hat\alpha\hat\mu\hat\nu\hat\beta}\theta\right)
\left(\bar\lambda\gamma_{\hat\alpha\hat\mu\hat\nu\hat\psi}\lambda\right).}
This term lacks a manifest factor of $(ss^\prime-1)$ and
hence appears to mediate an interaction between branes having the same
sign charge, as well as between branes having opposite charges. Such
an interaction would not be entirely surprising, or inconsistent.  We
can expect an exact force cancellation between branes only if their
spins combine to form an overall BPS state. Since there are $256$ spin
states in a BPS multiplet, we would expect that $256\times 255$ of the
possible combinations of spins of two BPS objects would experience a
net force. However, it turns out in the present case that $L_3$ does
vanish for $ss^\prime=1$ as a consequence of
the projection conditions \fzero\ and \thetafzero\ satisfied by $\lambda$ and $\theta$.
These conditions can be used to show that
\eqn\identity{A^{\hat\beta\hat\psi}\equiv\ 
\left(\bar\theta\gamma^{\hat\alpha\hat\mu\hat\nu\hat\beta}\theta\right)
\left(\bar\lambda\gamma_{\hat\alpha\hat\mu\hat\nu}^{\ \ \ \ \hat\psi}\lambda\right)
= \ {ss^\prime\over 4}\delta^{\hat\beta\hat\psi}
\left(\bar\theta\gamma^{\hat\sigma_1\dots\hat\sigma_4}\theta\right)
\left(\bar\lambda\gamma_{\hat\sigma_1\dots\hat\sigma_4}\lambda\right)
\ -ss^\prime A^{\hat\psi\hat\beta}.}
This shows that for $ss^\prime=1$, the symmetric part of the matrix
$A^{\hat\beta\hat\psi}$ is proportional to the identity.  The
prefactor in
\lthree\ is symmetric, so contracting with the antisymmetric part of 
$A^{\hat\beta\hat\psi}$ gives zero. The remaining term is proportional
to the Laplacian of $f$ which vanishes in the long distance limit.  Note that for $ss^\prime=-1$,
there will still be a non-zero interaction, which is proportional to
the product of the dipole moment $\mu_{\alpha\beta\rho\sigma}$ 
\seconddipole\ 
of the background superpartner and the corresponding dipole moment of
the probe brane.

\newsec{Conclusion}
\noindent
In this article we have constructed the spinning superpartner
spacetimes that fill out the BPS multiplet of membrane solitons of
M-theory.  The construction involved quantizing the fermionic
zero-modes of the soliton and resulted in operator valued spacetime
fields that acquired a conventional classical meaning only after taking
expectation values in fixed BPS states.  We examined the interaction
of spinning branes by studying the effective action of a stationary
spinning probe placed in the superpartner spacetimes.  Amongst other
interesting effects we found gravitational spin-spin interactions and
fermion-number changing gravitino exchanges.  We have shown that all
the interaction terms in \final\ vanish between two like-charge
branes, independently of their spin states.  This does not imply that
all static forces cancel between like-charge objects -- indeed, only
specific choices of the relative spin state should preserve the BPS
property.  Nevertheless, our results are consistent with recent
work on the spin dependent interactions of D$0$-branes
\refs{\harveyspin,\krausspin,\morales,\barrio}, 
where the non-zero static interactions begin at eighth order in
fermionic parameters, beyond the fourth order terms computed here.

\bigskip
{\bf Acknowledgements: } 
We would like to thank Jon Bagger and John Donoghue for useful discussions. 
The work of D.K., J.T. and K.Z.W. is supported in part by NSF grant NSF-THY-8714-684-A01.
V.B. would like to thank the University of
Massachusetts at Amherst, the Aspen Center for Physics, and Rutgers
University for hospitality during various stages of this work.
V.B. is supported by the Harvard Society of Fellows and by NSF grant
NSF-PHY-9802709.

\listrefs
\bye